\documentclass[aps,pre,twocolumn,superscriptaddress,nofootinbib,longbibliography]{revtex4-2}
\usepackage{amsmath,amssymb}
\usepackage{graphicx}
\usepackage[colorlinks=true,linkcolor=blue,citecolor=blue,urlcolor=blue]{hyperref}

\newcommand{\Ex}{\mathbb{E}}
\newcommand{\Tm}{\mathcal{T}}
\newcommand{\Jc}{\mathcal{J}}
\newcommand{\rv}{\mathbf{r}}
\newcommand{\uv}{\mathbf{u}}

\begin{document}

\title{Optimal probing scale for current fluctuations\\ in a Brownian gyrator}

\author{Badr Farih}
\email{badr.farih1@usmba.ac.ma}


\begin{abstract}
A driven colloidal rotor sustains a circulating current whose fluctuations one
would like to bound. Any such bound rests on how strongly the current responds
to a perturbation and on how much extra dissipation that perturbation costs, so
there is a well-posed question of \emph{where} to push: the response per unit
Onsager--Machlup cost depends on the radius at which the probe acts, and is
maximised at a definite one. We answer this for the Brownian gyrator and a
quadrupolar shear under a Gaussian envelope, a probe chosen so that linear
response is exactly blind to it at every observation window, so that the entire
signal is second order. The problem separates: the cost is radial and does not
see the circulation, while the response lives in the $m=3$ angular sector, where
the resolvent reduces to Kummer's equation with $b=4$ and the susceptibility is
a hypergeometric function of the envelope width. Maximising the ratio gives the
optimal probing radius in closed form. It is set by whichever of the system's two
clocks is faster: for weak driving $r^{*}=1.1264\sqrt{D/\gamma}$, the thermal
radius of the trap, and for strong driving $r^{*}=1.5563\sqrt{D/\Omega}$, the
distance diffused in one radian of rotation, with the trap stiffness dropping out
entirely. The crossover is at $\Omega=\gamma$. Both limits, and the prefactors,
are confirmed against direct simulation.
\end{abstract}

\maketitle

\section{Introduction}

The Brownian gyrator~\cite{Filliger2007,Argun2017,Chang2021} is the simplest
model of a driven colloidal rotor: a particle in a harmonic trap subject to a
non-conservative circulating force. It settles into a steady state that carries
a persistent angular current and produces entropy at a constant rate, and it has
been realised experimentally with optical tweezers and with electrical
analogues~\cite{Argun2017,Chiang2017}.

Suppose one wants to bound the fluctuations of that current. Every such bound,
from Cram\'er--Rao through the thermodynamic uncertainty
relations~\cite{Barato2015,Gingrich2016,Seifert2012}, has the same shape: a
response in the numerator and a cost in the denominator. Perturb the system,
measure how much the current moves, divide by how much the perturbation
dissipates. Nothing in that recipe fixes \emph{where in space} the perturbation
should act, and the answer is not indifferent: a probe concentrated deep inside
the trap barely moves the current, while one spread far outside it costs a great
deal of dissipation for little return. There is an optimal radius, and this paper
computes it.

The answer turns out to be governed by a competition between the system's two
time scales. The trap relaxes at rate $\gamma$ and the drive rotates at rate
$\Omega$, and each defines a length through the diffusion constant:
$\sqrt{D/\gamma}$, the thermal radius, and $\sqrt{D/\Omega}$, the distance
diffused in one radian of rotation. We find that the optimal probing radius
locks onto whichever is smaller,
\begin{equation}
r^{*}=
\begin{cases}
1.1264\,\sqrt{D/\gamma}, & \Omega\ll\gamma,\\[4pt]
1.5563\,\sqrt{D/\Omega}, & \Omega\gg\gamma,
\end{cases}
\label{eq:headline}
\end{equation}
with the crossover at $\Omega=\gamma$ and with the trap stiffness dropping out
of the strongly driven case altogether. The prefactors are not fitted: the
second is $\sqrt{2/c_0}$ with $c_0$ the root of an equation in sine and cosine
integrals, derived in Sec.~\ref{sec:scale}.

Two features of the construction make this computable. First, we choose a probe
to which linear response is exactly blind, so the leading signal is the
second-order susceptibility and there is no first-order term to subtract; this is
a design choice, not an obstruction, and Sec.~\ref{sec:probe} shows the blindness
holds at every finite observation window rather than only in the stationary
limit. Second, the resulting problem separates cleanly into a radial sector,
which turns out not to see the circulation at all, and a single angular sector,
in which the resolvent is a confluent hypergeometric equation. The optimisation
is then a one-dimensional variational problem with a closed-form objective.

\section{Model and observable}
\label{sec:model}

The gyrator is the two-dimensional linear diffusion
\begin{equation}
d\rv_t=\mathbf b_0(\rv_t)\,dt+\sqrt{2D}\,d\mathbf W_t,
\quad
\mathbf b_0=-\gamma\rv+\Omega(-y,x)^{\top},
\label{eq:sde}
\end{equation}
with isotropic diffusion. The circulation is orthogonal to $\nabla r^{2}$, so it
does not enter the radial balance and the steady state is the isotropic Gaussian
\begin{equation}
p_0(\rv)=\frac{1}{2\pi\sigma_0^{2}}e^{-r^{2}/2\sigma_0^{2}},
\qquad
\sigma_0^{2}=\frac{D}{\gamma}.
\label{eq:p0}
\end{equation}

The observable is the time-averaged circulating current,
\begin{equation}
\Jc=\frac{1}{\Tm}\int_0^{\Tm}\!\big(X\circ dY-Y\circ dX\big),
\label{eq:J}
\end{equation}
a Stratonovich integral whose deterministic part is $\Omega r^{2}$, so that
\begin{equation}
\langle\Jc\rangle_0=\Omega\,\Ex_0[r^{2}]=\frac{2\Omega D}{\gamma}.
\label{eq:J0}
\end{equation}
Being odd under time reversal, a non-zero $\langle\Jc\rangle_0$ marks sustained
breaking of detailed balance. Direct simulation of Eq.~\eqref{eq:sde} at
$\gamma=1.2$, $D=0.5$, $\Omega=0.9$, $\Tm=6$ on $2\times10^{5}$ trajectories
gives $\langle\Jc\rangle_0=0.751525$ against the predicted $0.750000$, a
discrepancy of $1.5$ standard errors.

One structural fact is used repeatedly below. Since
$\rv\cdot\Omega(-y,x)^{\top}=0$, the radial coordinate $r_t$ evolves under
$\gamma$ and $D$ alone: \emph{the radial process does not see the circulation}.
Every purely radial quantity below is therefore $\Omega$-independent, and the
whole dependence of the final answer on the drive is carried by one angular
sector.

\section{A probe linear response cannot see}
\label{sec:probe}

Perturb the drift by $\mathbf b_\lambda=\mathbf b_0+\uv_\lambda$ with the
quadrupolar shear under a Gaussian envelope,
\begin{equation}
\uv_\lambda=\lambda\,g(r)\begin{pmatrix}2xy\\x^{2}-y^{2}\end{pmatrix},
\quad
g(r)=e^{-\alpha r^{2}/2\sigma_0^{2}},
\quad\alpha>0,
\label{eq:field}
\end{equation}
whose radial and azimuthal projections are
$\lambda r^{2}g\sin3\theta$ and $\lambda r^{2}g\cos3\theta$. The parameter
$\alpha$ sets the width of the envelope, and the field magnitude
$|\uv_\lambda|=|\lambda|r^{2}g(r)$ peaks at
\begin{equation}
r^{*}=\sigma_0\sqrt{2/\alpha},
\label{eq:rstar}
\end{equation}
which is the radius the paper is about.

Two properties make Eq.~\eqref{eq:field} usable as a probe.

\emph{It confines, for either sign.} Since
$\rv\cdot\mathbf b_\lambda=-\gamma r^{2}+\lambda r^{3}g\sin3\theta$ and
$rg(r)\to0$, the Lyapunov drift condition
$\limsup_{r\to\infty}(\rv\cdot\mathbf b_\lambda)/r^{2}=-\gamma<0$ holds for every
real $\lambda$ and every $\alpha>0$, so the perturbed process remains
non-explosive and exponentially ergodic~\cite{MeynTweedie1993}. Without the
envelope this fails for one sign: the bare polynomial field drives the particle
outward faster than the trap restores it, in the sectors where
$\lambda\sin3\theta>0$. The mollification buys two-sidedness, which matters in
Sec.~\ref{sec:cost}. Numerically, $\max|\rv|=3.767,3.765,3.762,3.761$ at
$\lambda=-0.8,-0.3,0.3,0.8$: no sign dependence, no divergence.

\emph{Linear response is blind to it, at every window.} The first-order response
of $\Jc$ has two sources, and both vanish by angular orthogonality. The
observable's own shift is $xu_{\lambda,y}-yu_{\lambda,x}=\lambda r^{3}g\cos3\theta$,
averaged at this order in the isotropic law Eq.~\eqref{eq:p0}. The density shift
obeys $\mathcal L_0^{\dagger}p_1=\nabla\!\cdot\!(\uv_\lambda p_0)$ with
\begin{equation}
\nabla\!\cdot\!(\uv_\lambda p_0)
=-\lambda(1+\alpha)\frac{r^{3}g}{\sigma_0^{2}}\,p_0\sin3\theta,
\label{eq:source}
\end{equation}
a pure $m=\pm3$ source; the unperturbed generator is rotationally covariant, so
$m$ is conserved at every instant and $p_1(t)$ never acquires an isotropic
component to pair with $\Omega r^{2}$. Hence $\chi_1(\Tm)=0$ for every $\Tm$,
not merely asymptotically. Because both signs of $\lambda$ are admissible this
can be checked by a central difference, which cancels the even part exactly: the
measured odd part sits at $0.002$, $0.004$ and $0.010$ standard errors for
$\lambda=0.2,0.4,0.8$.

The leading response is therefore second order,
$\Delta\langle\Jc\rangle=\tfrac12\chi_2\lambda^{2}+O(\lambda^{4})$, and $\chi_2$
is what the probe measures.

\section{Response per unit dissipation}
\label{sec:cost}

The cost of the perturbation is its Onsager--Machlup action. Writing
$M_\lambda$ for the Girsanov density of the tilt relative to the unperturbed
path measure~\cite{Liptser2001}, the relevant cross-moments are
$\Ex_0[M_\lambda M_\mu]=\Ex_{\lambda+\mu}[\exp\int\uv_\lambda\!\cdot\!\uv_\mu/2D]$,
and on this family the cross-potential collapses to a radial function,
\begin{equation}
\uv_\lambda\!\cdot\!\uv_\mu=\lambda\mu\,r^{4}g(r)^{2},
\qquad
Q\equiv\int_0^{\Tm}\!\frac{r_t^{4}g(r_t)^{2}}{2D}\,dt .
\label{eq:Q}
\end{equation}

For a pair of probes of equal and opposite amplitude the sum vanishes, so the
cross term is an average over the \emph{unperturbed} process. The resulting
$2\times2$ Gram matrix has the symmetric vector as an eigenvector, the response
being even in $\lambda$, and the bound is
$2\Delta\langle\Jc\rangle^{2}/(\chi^{2}+\rho)$ with
$\chi^{2}=\Ex[e^{\lambda^{2}Q}]-1$ and $\rho=\Ex_0[e^{-\lambda^{2}Q}]-1<0$.
Numerically this design is well conditioned at moderate amplitude
($\kappa\simeq6.4/\lambda^{2}$, of order ten at $\lambda\simeq0.8$), and adding
further probe amplitudes gains nothing: the best designs at three, four and five
amplitudes exceed the symmetric pair by $0.1\%$ while their condition numbers
reach $1.2\times10^{4}$, $4.3\times10^{6}$ and $1.7\times10^{8}$, which is
numerical noise rather than signal.

The bound also decreases monotonically in $\lambda$, so its supremum is the
zero-amplitude limit. Both the numerator and $\chi^{2}+\rho$ are $O(\lambda^{4})$
there, the latter because the symmetric pairing cancels the $O(\lambda^{2})$
part of the Gram, and the ratio tends to
\begin{equation}
\mathcal B=\frac{\chi_2^{2}}{2\,\Ex_0[Q^{2}]},
\label{eq:B}
\end{equation}
which is the second-order Bhattacharyya bound~\cite{Bhatt1946,Barankin1949} for
this family. We use Eq.~\eqref{eq:B} as the figure of merit throughout: it is
response squared per unit Onsager--Machlup cost, and the question of the paper is
which envelope width maximises it.

\section{Exact solution}
\label{sec:exact}

\subsection{The cost is radial}

By the remark of Sec.~\ref{sec:model} the radial process is
$\Omega$-independent, so $\Ex_0[Q^{2}]$ does not depend on the drive. Measured
at $\alpha=1$ it comes out $9.6915\times10^{-2}$, $9.6921\times10^{-2}$ and
$9.7065\times10^{-2}$ at $\Omega/\gamma=0.25,0.75,2$. In the stationary regime
$\Ex_0[Q^{2}]\simeq(\Tm I_Q/2D)^{2}$ with
\begin{equation}
I_Q(\alpha)=\Ex_0\!\big[r^{4}e^{-\alpha r^{2}/\sigma_0^{2}}\big]
=\frac{8\sigma_0^{4}}{(1+2\alpha)^{3}},
\label{eq:IQ}
\end{equation}
exact and reproduced by quadrature to machine precision. Simulation gives
$\Ex_0[Q^{2}]/(\Tm I_Q/2D)^{2}=1.145,1.056,1.018,1.028,1.074$ at
$\alpha=\tfrac14,\tfrac12,1,2,4$, the excess being the variance of $Q$.

\subsection{The response is one angular sector}

The second-order response has two channels. Besides the observable's own shift,
the shear redistributes probability radially, and since
$\langle\Jc\rangle_0=\Omega\langle r^{2}\rangle$ this feeds back into the
current: It\^o's formula gives
$d\langle r^{2}\rangle/dt=-2\gamma\langle r^{2}\rangle+4D
+2\lambda\langle r^{3}g\sin3\theta\rangle$, so
$\delta\langle r^{2}\rangle=(\lambda/\gamma)\langle r^{3}g\sin3\theta\rangle_{p_1}$.
The second channel is the larger of the two.

Both reduce to the same radial integral. Writing $p_1=p_0\phi$ and using that
the $L^{2}(p_0)$ adjoint is the generator of the time-reversed process, whose
drift $-\gamma\rv-\Omega(-y,x)^{\top}$ has the same radial part and reversed
circulation, the equation for $\phi=\mathrm{Re}[\Phi e^{3i\theta}]$ with
$\Phi=r^{3}\psi$ becomes, in $z=r^{2}/2\sigma_0^{2}$, Kummer's equation with a
source,
\begin{equation}
z\psi''+(4-z)\psi'-a\psi=\frac{i(1+\alpha)}{2D}e^{-\alpha z},
\quad
a=\tfrac32\Big(1+i\frac{\Omega}{\gamma}\Big).
\label{eq:kummer}
\end{equation}
The factor $r^{3}$ removes the centrifugal term and $z$ does the rest. Expanding
in generalised Laguerre polynomials $L_n^{(3)}$, which diagonalise the operator
on the left, and using
$\int_0^{\infty}z^{3}e^{-(1+\alpha)z}L_n^{(3)}dz=\Gamma(n{+}4)\alpha^{n}/
[n!(1+\alpha)^{n+4}]$ both to solve and to project onto the observable, gives
\begin{equation}
\chi_2(\alpha)=\frac{4\sigma_0^{6}}{D}
\Big[\mathrm{Im}\,\Sigma+\frac{\Omega}{\gamma}\,\mathrm{Re}\,\Sigma\Big],
\label{eq:chi2}
\end{equation}
\begin{equation}
\Sigma(\alpha)=\frac{6}{a(1+\alpha)^{7}}\,
{}_2F_1\!\Big(4,a;a+1;\Big(\frac{\alpha}{1+\alpha}\Big)^{2}\Big).
\label{eq:2f1}
\end{equation}
The hypergeometric form follows from $(4)_n/n!=(n{+}1)(n{+}2)(n{+}3)/6$ and
$(a)_n/(a{+}1)_n=a/(n{+}a)$; it agrees with the series to $10^{-17}$ over
$\alpha\in[0.5,20]$ and $\Omega\in[0.3,2.4]$.

A by-product of the same diagonalisation is the spectrum of the sector,
$\lambda_n=2\gamma(n+a)=(2n{+}3)\gamma+3i\Omega$. The slowest mode relaxes at
$\mathrm{Re}\,\lambda_0=3\gamma$, and $|\lambda_0|=3\sqrt{\gamma^{2}+\Omega^{2}}$
grows with the drive; this governs how quickly a finite observation window
approaches the stationary result, and is why the residuals in
Sec.~\ref{sec:test} shrink as $\Omega/\gamma$ rises.

Equation~\eqref{eq:chi2} is the stationary response, and simulation approaches
it from below as the window lengthens. At $\Omega/\gamma=3/4$ the measured
$\chi_2$ is $0.8267$, $0.9012$ and $0.9507$ of Eq.~\eqref{eq:chi2} at
$\Tm=6,12,24$ for $\alpha=1$, and $0.8770$, $0.9281$, $0.9639$ for $\alpha=2$.
The deficit halves as the window doubles; Richardson extrapolation in $1/\Tm$
gives $1.0002$ and $0.9998$.

\section{The optimal probing scale}
\label{sec:scale}

\begin{figure}[t]
\centering
\includegraphics[width=\columnwidth]{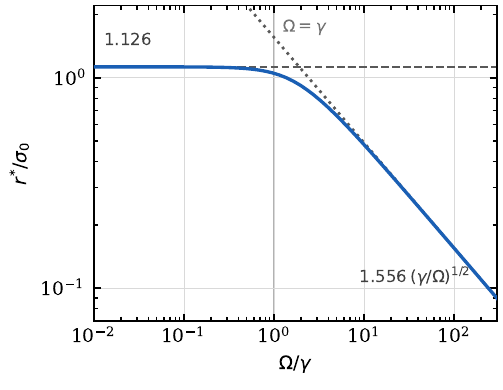}
\caption{The optimal probing radius, from the maximiser of
Eq.~\eqref{eq:objective}. It plateaus at the thermal radius when the drive is slow
and contracts as $(\gamma/\Omega)^{1/2}$ when the drive is fast, crossing over
at $\Omega=\gamma$ (vertical line). Dashed and dotted lines are the two
asymptotes of Eq.~\eqref{eq:headline}; both prefactors are computed, not fitted.}
\label{fig:scale}
\end{figure}

Combining Eqs.~\eqref{eq:B}, \eqref{eq:IQ} and \eqref{eq:chi2} and dropping
$\alpha$-independent factors, the figure of merit is
\begin{equation}
\mathcal B(\alpha)\;\propto\;\big[\chi_2(\alpha)\,(1+2\alpha)^{3}\big]^{2},
\label{eq:objective}
\end{equation}
a one-dimensional variational problem whose maximiser fixes $r^{*}$ through
Eq.~\eqref{eq:rstar}. Figure~\ref{fig:scale} shows the result.

\emph{Weak driving.} As $\Omega/\gamma\to0$ the optimum saturates at
$\alpha^{*}\to1.57634$, already to four decimal places at $\Omega/\gamma=0.02$,
so $r^{*}\to1.12639\,\sigma_0=1.1264\sqrt{D/\gamma}$. Both
$\langle\Jc\rangle_0$ and $\chi_2$ vanish linearly in $\Omega$ in this limit,
but their ratio in Eq.~\eqref{eq:objective} does not, and the optimal width stays
finite. The scale is the trap's.

\emph{Strong driving.} Here $\alpha^{*}$ becomes proportional to the drive, and
the coefficient can be computed. Setting $\alpha=cq$ with $q=\Omega/\gamma\to
\infty$, the Laguerre sum is dominated by $n=O(\alpha)$; substituting
$n=\alpha u$ turns it into an integral, the leading $O(q^{-1})$ terms of
$\mathrm{Im}\,\Sigma+q\,\mathrm{Re}\,\Sigma$ cancel, and Eq.~\eqref{eq:objective}
tends to a function of $c$ alone,
\begin{equation}
\mathcal B\to\big[\tfrac23\beta\,\mathcal I(\beta)\big]^{2},
\quad
\beta=\frac{3}{2c},
\quad
\mathcal I=\int_0^{\infty}\!\frac{u^{4}e^{-2u}}{u^{2}+\beta^{2}}du .
\label{eq:limit}
\end{equation}
The integral is elementary,
\begin{equation}
\mathcal I(\beta)=\tfrac14-\tfrac{\beta^{2}}{2}
+\beta^{3}\big[\mathrm{Ci}(2\beta)\sin2\beta
+(\tfrac{\pi}{2}-\mathrm{Si}(2\beta))\cos2\beta\big],
\label{eq:Ibeta}
\end{equation}
and maximising gives $\beta_0=1.816493$, hence $c_0=3/2\beta_0=0.825767$. Then
$\alpha^{*}\simeq c_0\,\Omega/\gamma$ and, from Eq.~\eqref{eq:rstar},
\begin{equation}
r^{*}\simeq\sigma_0\sqrt{\frac{2}{c_0}}\Big(\frac{\gamma}{\Omega}\Big)^{1/2}
=\sqrt{\frac{2}{c_0}}\,\sqrt{\frac{D}{\Omega}}
=1.5563\sqrt{\frac{D}{\Omega}} .
\label{eq:strong}
\end{equation}
The trap stiffness has cancelled. What remains is the distance the particle
diffuses in one radian of rotation, and the probe should sit there. Numerically
$\alpha^{*}\gamma/\Omega=0.8752,0.8456,0.8343,0.8297,0.8276$ at
$\Omega/\gamma=8,16,32,64,128$, converging on $c_0$; and varying $\gamma$ over
an order of magnitude at fixed $\Omega$ leaves $r^{*}\sqrt{\Omega/D}$ at
$1.5514,1.5548,1.5423,1.5540,1.5552$, confirming that $\gamma$ has genuinely
dropped out rather than been absorbed.

The square-root law is an asymptote, not a global scaling. Across the crossover
$r^{*}$ varies only between $1.13\,\sigma_0$ and $0.92\,\sigma_0$ while
$\Omega/\gamma$ runs over two decades, so a power law fitted through that region
would describe neither limit.

\section{Test}
\label{sec:test}

The prediction that $r^{*}$ tracks the drive is falsifiable, and because
$\Ex_0[Q^{2}]$ is $\Omega$-independent the entire effect must come from
$\chi_2$. Table~\ref{tab:test} compares the predicted optimum against the peak
of the measured $\mathcal B$ curve.

\begin{table}[b]
\caption{Optimal envelope width, predicted against measured. The measured value
locates the peak of $\mathcal B$ by a parabolic fit on an $\alpha$ grid at
$\Tm=6$, $1.2\times10^{5}$ trajectories.}
\label{tab:test}
\begin{ruledtabular}
\begin{tabular}{lccc}
$\Omega/\gamma$ & theory ($\Tm\to\infty$) & measured ($\Tm=6$) & gap \\
\hline
0.25 & 1.592 & 1.843 & $+0.251$ \\
0.75 & 1.715 & 1.867 & $+0.152$ \\
2.00 & 2.382 & 2.412 & $+0.030$ \\
\end{tabular}
\end{ruledtabular}
\end{table}

Both sequences rise, and the residual behaves as the spectrum of
Sec.~\ref{sec:exact} requires: the finite-window deficit is controlled by
$|\lambda_0|=3\sqrt{\gamma^{2}+\Omega^{2}}$, so it shrinks as the drive
stiffens, and at $\Omega/\gamma=2$ theory and measurement agree to $1.3\%$. The
slowest case confirms the diagnosis directly: at $\Omega/\gamma=0.25$ on a fixed
grid the measured optimum moves from $1.780$ at $\Tm=6$ to $1.697$ at
$\Tm=18$, against $1.592$ predicted. A single mode accounts for the direction
and the ordering but not the magnitude, the measured deficit exceeding the
one-mode estimate by a factor of $2.6$ to $3.6$, since the source projects onto
every $n$ and the time average carries its own transient.

\section{Discussion}

The question of where to probe a driven system has a definite answer, and for
the gyrator it is set by a competition of time scales rather than by the
geometry of the trap. When the drive is slow the optimal radius is the thermal
one, $\sqrt{D/\gamma}$; when the drive is fast it is $\sqrt{D/\Omega}$, and the
stiffness that defines the trap plays no role at all. That the two regimes meet
at $\Omega=\gamma$ is unsurprising; that the strongly driven scale forgets
$\gamma$ entirely is less so, and it is the cleanest statement the calculation
produces.

The structure that makes this tractable is the separation of the problem into a
radial sector that cannot see the circulation and one angular sector that
carries all of it. We would expect the separation, though not the particular
constants, to survive for other probe symmetries: the choice of the $m=3$
quadrupolar shear fixes which Laguerre family appears and hence the $b=4$ in
Eq.~\eqref{eq:kummer}, but the argument that the radial cost is $\Omega$-blind uses
only $\rv\cdot\Omega(-y,x)^{\top}=0$.

Several things are left open. We have optimised the width of a fixed angular
profile; whether a different mode or radial shape yields a larger $\chi_2$ in
the first place is untested, and would change the prefactors in
Eq.~\eqref{eq:headline} without, we expect, changing the two scaling regimes. The
observation window has been treated as a nuisance to be extrapolated away rather
than as a variable of interest, though the spectrum $\lambda_n=2\gamma(n+a)$
suggests the finite-$\Tm$ problem is equally tractable. And the experimental
question is open: in an optical-tweezer realisation~\cite{Argun2017} both
$\gamma$ and $\Omega$ are tunable, and the crossover in Fig.~\ref{fig:scale}
occurs at accessible values.

\begin{acknowledgments}
I thank my mom for unplugging the Wi-Fi that night; it seems ideas come when
you have nothing else to do.
\end{acknowledgments}

\section*{Data availability}

No new experimental data were created. All reported quantities are closed-form
expressions or the output of the simulation protocol described in the text. The
scripts generating every number and figure are deposited at Zenodo,
DOI:10.5281/zenodo.22877097. Diffusions are integrated by Euler--Maruyama,
Stratonovich integrals by the midpoint rule, and every response is differenced
across amplitudes on shared Brownian increments including the unperturbed leg.
All scripts are seeded, and independent reruns reproduce every printed value exactly.


\begin{thebibliography}{99}

\bibitem{Filliger2007}
R.~Filliger and P.~Reimann,
Phys. Rev. Lett. \textbf{99}, 230602 (2007).

\bibitem{Argun2017}
A.~Argun, J.~Soni, L.~Dabelow, S.~Bo, G.~Pesce, R.~Eichhorn, and G.~Volpe,
Phys. Rev. E \textbf{96}, 052106 (2017).

\bibitem{Chang2021}
H.~Chang, C.-L.~Lee, P.-Y.~Lai, and Y.-F.~Chen,
Phys. Rev. E \textbf{103}, 022128 (2021).

\bibitem{Chiang2017}
K.-H.~Chiang, C.-L.~Lee, P.-Y.~Lai, and Y.-F.~Chen,
Phys. Rev. E \textbf{96}, 032123 (2017).

\bibitem{Barato2015}
A.~C.~Barato and U.~Seifert,
Phys. Rev. Lett. \textbf{114}, 158101 (2015).

\bibitem{Gingrich2016}
T.~R.~Gingrich, J.~M.~Horowitz, N.~Perunov, and J.~L.~England,
Phys. Rev. Lett. \textbf{116}, 120601 (2016).

\bibitem{Seifert2012}
U.~Seifert,
Rep. Prog. Phys. \textbf{75}, 126001 (2012).

\bibitem{MeynTweedie1993}
S.~P.~Meyn and R.~L.~Tweedie,
Adv. Appl. Probab. \textbf{25}, 518 (1993).

\bibitem{Liptser2001}
R.~S.~Liptser and A.~N.~Shiryaev,
\emph{Statistics of Random Processes I}, 2nd ed.
(Springer, Berlin, 2001).

\bibitem{Bhatt1946}
A.~Bhattacharyya,
Sankhy\=a \textbf{8}, 1 (1946).

\bibitem{Barankin1949}
E.~W.~Barankin,
Ann. Math. Statist. \textbf{20}, 477 (1949).


\end{thebibliography}
\end{document}